\documentclass[amsmath,amssymb,twocolumn, longbibliography]{revtex4-2}

\usepackage[colorlinks,bookmarks=true,citecolor=blue,linkcolor=red,urlcolor=blue]{hyperref}
\usepackage{graphicx}
\usepackage{subfigure}
\usepackage{adjustbox}
\usepackage{bm}
\usepackage{color}
\usepackage{braket}
\usepackage{standalone}
\usepackage{multirow}
\usepackage{tikz}
\usepackage{dsfont}

\newcommand{\bea}{\begin{eqnarray}}
\newcommand{\eea}{\end{eqnarray}}
\definecolor{dgreen}{rgb}{0.1,0.5,0.1}
\definecolor{babyblue}{rgb}{0.54, 0.81, 0.94}

\begin{document}

\title{Nodal Spectral Functions Stabilized by Non-Hermitian Topology of Quasiparticles}
\author{Carl Lehmann$^1$}
\email{carl.lehmann1@tu-dresden.de}
\author{Tommaso Micallo$^1$}
\author{Jan Carl Budich$^{1,2}$}

\affiliation{$^1$Institute of Theoretical Physics${\rm ,}$ Technische Universit\"{a}t Dresden and W\"{u}rzburg-Dresden Cluster of Excellence ct.qmat${\rm ,}$ 01062 Dresden${\rm ,}$ Germany}
\affiliation{$^2$Max Planck Institute for the Physics of Complex Systems${\rm ,}$ N\"othnitzer Str.~38${\rm ,}$ 01187 Dresden${\rm ,}$ Germany}

\date{\today}

\begin{abstract}
In quantum materials, basic observables such as spectral functions and susceptibilities are determined by Green's functions and their complex quasiparticle spectrum rather than by bare electrons. Even in closed many-body systems, this makes a description in terms of effective non-Hermitian (NH) Bloch Hamiltonians natural and intuitive. Here, we discuss how the abundance and stability of nodal phases is drastically affected by NH topology. While previous work has mostly considered complex degeneracies known as exceptional points as the NH counterpart of nodal points, we propose to relax this assumption by only requiring a crossing of the real part of the complex quasiparticle spectra, which entails a band crossing in the spectral function, i.e. a nodal spectral function. Interestingly, such real crossings are topologically protected by the braiding properties of the complex Bloch bands, and thus generically occur already in one-dimensional systems without symmetry or fine-tuning. We propose and study a microscopic lattice model in which a sublattice-dependent interaction stabilizes nodal spectral functions. Besides the gapless spectrum, we identify non-reciprocal charge transport properties after a local potential quench as a key signature of non-trivial band braiding. Finally, in the limit of zero interaction on one of the sublattices, we find a perfectly ballistic unidirectional mode in a non-integrable environment, reminiscent of a chiral edge state known from quantum Hall phases. Our analysis is corroborated by numerical simulations both in the framework of exact diagonalization and within the conserving second Born approximation.

\end{abstract}

\maketitle

\section{ Introduction}
Level (anti-)crossings are of key importance in quantum physics, and have been subject to intense study for many decades, e.g. in the context of level statistics \cite{Wigner1955,Dyson1962}. In condensed matter physics, topological semimetals characterized by stable nodal points in their Bloch band structure represent a frontier of current research \cite{Wehling2014,Armitage2018,Hasan2010,Qi2011}. The advent of non-Hermitian (NH) topological phases \cite{Ashida2020,Yao2018,Kawabata2019,Shen2018,Bergholtz2019,Lee2019,Wang2021,Li2021,Rui2022,Hu2021,Wojcik2020} accounting for dissipative effects such as finite lifetime of quasiparticles has recently provided a new perspective on nodal band structures and their topological stability \cite{Kozii2017,Yoshida2018,Rausch2020,Michishita2020,Michishita22020,Nagai2020,Crippa2021,Peter2021}. In particular, degeneracies in the complex spectra of effective NH Hamiltonians generically occur in the non-diagonalizable form of exceptional points (EPs) \cite{Bergholtz2019,Zhang2012} which, quite remarkably, are more abundant than diagonalizable nodes in the Hermitian realm. As a consequence stable EPs are found in two spatial dimensions (2D) whereas their Hermitian counterpart known as Weyl points are robust only in 3D systems \cite{Armitage2018}. Hence, the onset of dissipation may stabilize fine-tuned or symmetry protected nodal points to topologically protected EPs.

In recent studies \cite{Miri2019,Berry2004,Michen2021,Okugawa2019,Budich2019}, EPs have been considered as the NH analog of band (or level) crossings in a wide range of physical settings, including condensed matter systems \cite{Kozii2017,Yoshida2018,Rausch2020,Michishita22020,Nagai2020,Crippa2021,Peter2021}. In the latter case, complex energies become relevant as complex poles of the single-particle Green's function (GF) $\hat{G}^R(k,\omega)\propto \frac{1}{\omega - e(k)}$, describing excitations in terms of a quasiparticle dispersion \cite{Stefanucci2010,Fabrizio2020,Woelfle2018,Kozii2017,Economou2006,Martin2016} $e(k)$.
However, we note that in basic physical observables such as spectral functions $A(k,\omega) = -\frac{1}{\pi}\text{Im}\text{Tr}(\hat{G}^R(k,\omega))$, the most decisive quantity is the real part of the quasiparticle dispersion $\text{Re}(e(k))$, whereas the imaginary part mostly determines how sharp or blurred the quasiparticle band appears in a measurement. We thus find it natural to define  {\textit{nodal spectral functions}} through the occurrence of crossings in the real part of the quasiparticle spectra, rather than requiring full complex degeneracies tantamount to EPs.

\begin{figure}[htp]
	\centering
	\includegraphics[width=1.0\linewidth]{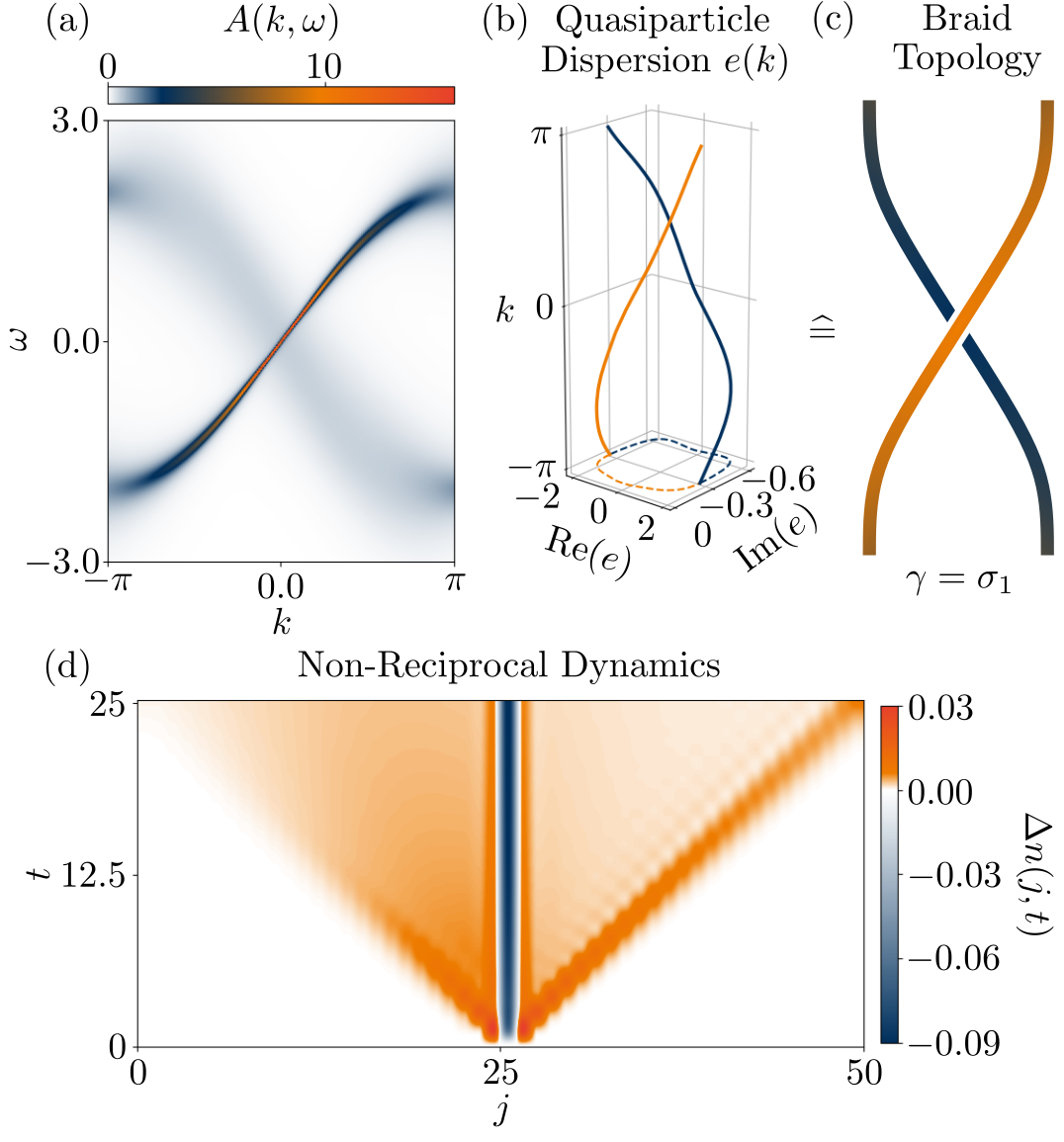}
	\caption{(a) Spectral function $A(k,\omega)$ of the model defined in Eq.~(\ref{EQ:Hamiltonian}) in a non-trivial braiding phase $c=1$, $\gamma=\sigma_1$ for periodic boundary conditions.  (b) Corresponding complex quasiparticle dispersion exhibiting non-trivial braiding, and (c) schematic link diagram. (d) Non-reciprocal propagation of a density excitation after switching on an onsite potential  in the middle of the system at site $j$. Parameters in all plots: $N=200$ unit cells, $J=1.0,\tau_0=1.0,\tau_1=0.0,v_0=-1.0, v_1=0.0, U_a=1.5,U_b=0.1$. }
	\label{fig:overview}
\end{figure}

Below, we investigate the stability of nodal spectral functions, and identify both equilibrium and dynamical phenomena characterizing the corresponding NH topological phases. Importantly, EPs here appear as the transition points between phases exhibiting protected crossings in the spectral functions, and gapped phases. The predicted nodal phases themselves are thus even more abundant than EPs and occur in 1D systems without requiring symmetry or fine-tuning. Their topological stability may be intuitively understood in terms of non-trivial braids \cite{Wang2021,Li2021,Rui2022,Hu2021,Wojcik2020,Nehra2022,Koenig2023} formed by the quasiparticle bands in the complex energy plane as a function of lattice momentum $k$: any non-trivial braid necessitates a crossing in the projection of the real part of the quasiparticle spectra, i.e. entails a nodal spectral function (see Fig.~\ref{fig:overview} (b-c)). 

We exemplify the occurrence of various topologically inequivalent spectral functions by solving a microscopic model of correlated electrons on a 1D lattice at finite temperature, both with exact diagonalization (ED) methods and within the conserving second Born approximation (SBA). Interestingly, if the imaginary part of the spectrum exhibits a large gap at the real crossing, such that one band is quite sharp while the other one is entirely blurred, the 1D nodal spectral function may arbitrarily closely mimic a chiral mode (cf.~Fig.~\ref{fig:overview} (a)), otherwise known only from edge states of 2D chiral topological phases \cite{Chiu2016}.
By studying the quench dynamics after a local perturbation in the particle density in such a pseudo-chiral scenario, we identify non-reciprocal transport signatures reminiscent of the uni-directional motion of a chiral edge state (see~Fig.~\ref{fig:overview} (d)). Our results underline that interacting many-body systems serve as a natural platform to observe interesting NH topological physics in basic measurable quantities such as spectral functions.

The remainder of this article is structured as follows:
The numerical methods and the microscopic lattice model are introduced in Section \ref{sec:MAM}, and the non-integrability of the model is studied in Section \ref{sec:NI}. In Section \ref{sec:ABG}, the notion of quasiparticle dispersions is defined and their classification using braid groups is discussed. In Section \ref{sec:CBT}, we detail the stabilizing effect on crossings (\ref{sec:TC}), the non-reciprocity following from the non-Hermitian character of the quasiparticle dispersion (\ref{sec:NRSPE}), the non-reciprocal effects within the charge-dynamics after a local quench (\ref{sec:NRCP}), and offer an intuitive explanation of the pseudo-chiral mode in terms of long-lived excitations on the weakly interacting sublattice (\ref{sec:LEUQ}). A concluding discussion is presented in Section \ref{sec:CON}.     

\section{\label{sec:MAM} Methods and Model}
Our main quantity of interest is the retarded GF describing the propagation of single-particle type excitations:
\begin{equation}
	\label{EQ:green_realtime}
	G^\mathrm{R}_{m,n}(t)=-\mathrm{i}\theta(t)\left< \{ c_m(0), c^\dagger_n (t) \} \right> \mathrm{,}
\end{equation}
at finite temperature $T=\frac{1}{\beta}$. Here $c_m(t) \left(c^\dagger_m(t) \right)$ denotes the annihilation(creation)-operator in the Heisenberg-picture acting on lattice-site $m$ and $\langle ...\rangle = \text{Tr}(\frac{1}{Z}\text{exp}(-\beta H)...)$ the expectation value with respect to the thermal Gibbs state, both involving the full Hermitian many-body Hamiltonian $H$. 
The GF is naturally non-Hermitian, due to interaction effects including inter-particle scattering. For stationary states such as thermal states, after Fourier-transformation in space and time, the GF can be rewritten by using the Dyson-equation \cite{Stefanucci2010} as
\begin{align}
	\label{eq:green1}
	\hat{G}^R(k,\omega) = \left(\omega \mathds{1} - \hat{h}(k) -\hat{\Sigma}(k,\omega)\right)^{-1} \text{.}
\end{align}
The GF is fully determined by the NH effective
Hamiltonian (eH) $\hat{h}_{e}(k,\omega) = \hat{h}(k) -\hat{\Sigma}(k,\omega)$, which consists of the Hermitian non-interacting Bloch-Hamiltonian $\hat{h}(k)$ and the NH self-energy $\hat{\Sigma}(k,\omega)$. The latter captures the scattering effects induced by interactions.
Since the Hermitian many-body Hamiltonian $H$ is interacting (see Section \ref{sec:MLM}) and non-integrable (see Section \ref{sec:NI}), extensive numerics is required in order to actually compute the GF and verify our analytical insights.
Our numerical results are based on two complementary methods: exact diagonalization (ED) and Non-Equilibrium Green's Functions (NEGFs) \cite{Stefanucci2010,Balzer2013,Aoki2014,Schueler2020}.
\paragraph*{Exact Diagonalization(ED)}
Within the ED framework we obtain exact results for small system sizes. Based on Eq.~(\ref{EQ:green_realtime}), the full unitary time-evolution of the thermal state after the action of an annihilation(creation)-operator is computed, leading to the GF. Our implementations are based on the python libraries NumPy and SciPy.
\paragraph*{Non-Equilibrium Green's Functions (NEGFs)}
The NEGF approach \cite{Stefanucci2010,Balzer2013,Aoki2014,Schueler2020}, as a (self-consistent) perturbative method in the interaction strength, reduces complexity to quadratic in system size and cubic in time. It allows, at moderate interaction strengths and not too low temperatures, reliable results (inline with ED) on large systems and enables to reach long time scales.
Here, the Kadanoff-Baym\cite{Stefanucci2010,Schueler2020} equations are solved numerically in conserving second Born approximation (SBA). The implementation is based on the software package NESSi\cite{Schueler2020}.
\subsection{\label{sec:MLM} Microscopic Lattice Model}

Inspired by previous works revealing non-trivial NH physics in related systems \cite{Ashida2020,Crippa2021,Kozii2017,Lehmann2021,Michishita2020,Nagai2020,Peter2021,Yoshida2018,Rausch2020}, a 1D interacting two-band model will be investigated, consisting of $N$ unit cells with two sublattices $A$ and $B$. The Hamiltonian $H$ and the interaction $\mathcal{V}$ can be written as:

\begin{align}
	\label{EQ:Hamiltonian}
	H &=\sum_k  c^\dagger_k \hat{h}(k) c_k  + \mathcal{V} \nonumber\\
	\mathcal{V} &=\sum_\alpha \sum_{j}U_ \alpha \left( n_{j, \alpha}-\frac{1}{2}\right)\left( n_{j+1, \alpha}-\frac{1}{2}\right) \text{ .}
\end{align}

Here $c_k=(c_{k, \mathrm{A}}, c_{k, \mathrm{B}})^T$ and $c^\dagger_k=(c^\dagger_{k, \mathrm{A}}, c^\dagger_{k, \mathrm{B}})$ are respectively the annihilation and creation operators in momentum-space acting on the two sublattice degrees of freedom. The matrix $\hat{h}(k)$ denotes the $2\times 2$ Bloch-Hamiltonian of the non-interacting system. Throughout this work we will use $\hat{h}(k) = (J-\tau_0 \cos(k) - \tau_1 \cos(2k))\hat{\sigma}_x + (v_0 \sin(k) + v_1 \sin(2k))\hat{\sigma}_z$ written in terms of Pauli-matrices $\{\hat{\sigma}_{\beta}\}$, describing nearest and next-nearest neighbour hoppings.
Further, the interaction $\mathcal{V}$ is given in real-space using the number operator $n_{j, \alpha}= c^\dagger_{j, \alpha} c_{j, \alpha}$ on site $j$ and sublattice $\alpha\in\{A,B\}$, where $c_{j, \alpha} =\frac{1}{\sqrt{N}} \sum_k e^{-ikj} c_{k, \alpha} $ defines the annihilation(creation) operator in real-space. We choose moderate interaction strengths, comparable but smaller than the non-interacting bandwidths, which also act dominantly on the $A$-sublattice $U_A>U_B$. 
All simulations are done in thermal equilibrium at fixed temperature $T=5$.
  
\subsection{\label{sec:NI}Non-Integrability Despite Free Sublattice}
\begin{figure}[htp]	 
	\centering 
	 
	\includegraphics[width=.9\columnwidth]{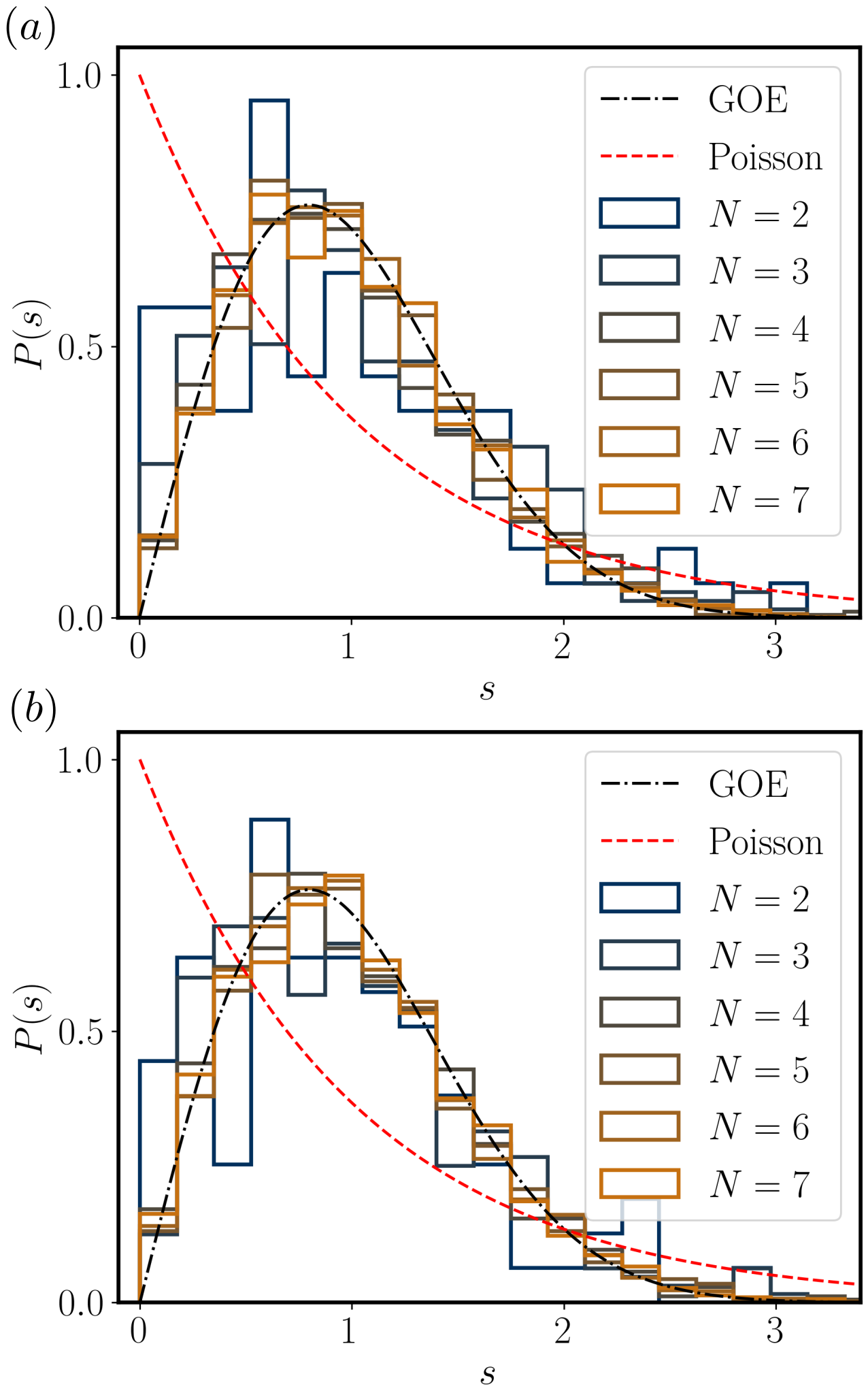}
	
	\caption{\label{fig:level_statistics}Level-spacing statistics for weakly interacting (panel (a), $U_B=0.1$) and non-interacting (panel (b), $U_B=0$) B-sublattice for a system of $L=14$ sites; other parameters in both panels: $J=1.0,\tau_0=1.0,\tau_1=0.0,v_0=-1.0, v_1=0.0$,  $U_A=1.5$, OBC. 
		The different lines refer to different particle numbers $N$, and at $N=7$ (half filling) the data refers to the particle-hole symmetrized sub-block.
		Both situations show a clear GOE behavior. Statistical fluctuations are bigger for lower $N$, where fewer states contribute.
		Results for $N>7$ are implied by particle-hole symmetry.}
\end{figure}

In the limit $U_B=0$, in our model system an exact single-particle excitation emerges which obeys coherent dynamics, i.e. exhibits infinite quasiparticle lifetime. We thus find it interesting to investigate for our non-standard 1D Hubbard model to investigate whether the full many-body Hamiltonian is non-integrable in the parameter regime of our present work. Interestingly, even in the regime of a non-interacting sublattice, we found a clear signature of chaotic behaviour in the level statistics, implying non-integrability.
To study the integrability of quantum systems, it is helpful to use spectral signatures, in particular fluctuations of energy level spacings \cite{Poilblanc1993,Relano2002,Santos2010}.
Via exact diagonalization of a finite-size system, it is possible to extract the full spectrum $\{E_i\}$ of the interacting many-body Hamiltonian.
Then, via the unfolding procedure~\cite{Relano2002}, the energies $E_{i}$ are mapped to dimensionless energies $e_i$, and the mean value of the renormalised energy spacings $s_i=e_{i+1}-e_i$ is $\left< s_i\right>=1$.
If the classical limit of the system under consideration is fully integrable, the level spacings $s_i$ show a Poisson statistics $P(s)=\mathrm{e}^{-s}$~\cite{Berry1977}.
If, instead, the classical limit of the quantum system is fully chaotic, the BGS conjecture states that Random Matrix Theory describes the spectral-spacing statistics~\cite{Bohigas1984}.
In particular, with symmetric models such as the one under study here (particle-hole symmetric), the system is expected to show the level spacing statistics associated to the Gaussian orthogonal ensemble (GOE) of random matrices, namely $P(s)=\frac{\pi\, s}{2}\mathrm{e}^{-\pi s^2/ 4}$.
In order to unfold the spectrum, all the symmetry sectors must be handled independently~\cite{Santos2010}.
Since the Hamiltonian~\eqref{EQ:Hamiltonian} is both particle-hole symmetric and particle-number conserving, the symmetry sectors corresponding to different filling fractions must be separated, and the block corresponding to half filling must be considered either in its symmetrised or anti-symmetrized sub-block with respect to particles and holes.
The results of this analysis are summarized in Fig.~\ref{fig:level_statistics}, showing a clear GOE distribution for the level spacings, both for a weakly-interacting and a non-interacting B sublattice, confirming that a strong interaction on one sublattice is enough to drive the system into a fully chaotic behaviour. 
Fig.~\ref{fig:level_statistics} shows data on systems with open boundary conditions (OBC) rather than periodic boundary conditions (PBC) because the absence of translational symmetry allows for a bigger number of states per symmetry sectors, leading to a higher statistical significance.

\section{\label{sec:ABG} Analytical Braiding-Topology of Electronic Band Structures } 
The rGF is typically understood in terms of quasiparticles \cite{Stefanucci2010,Woelfle2018,Fabrizio2020}, where as proposed by \cite{Economou2006,Martin2016,Kozii2017} the set of quasiparticle excitations $\{e_j(k)\}$ is formally defined by the set of poles of the rGF $\hat{G}^R(k,z) \propto \frac{1}{z-e_j(k)}$, allowing complex frequencies $z$. Since the rGF is here matrix-valued, we use the poles of the trace $\text{Tr}(\hat{G}^R(k,z))$. Note that the imaginary-part of this quantity on the real-frequency-axis corresponds to the well known spectral function. 
We stress that quasiparticles by definition are described by a generically complex dispersion, which represents a key property for our present work \footnote{It might be worth to note that one expects an infinite amount of complex poles, however in reality only the poles close to the real axis are of interest because they describe excitations on intermediate and long timescales.}.
In practice, when analyzing numerical data given only on the real-frequency-axis $\omega$, the (crude) assumption of a selfenergy that is constant with respect to imaginary frequency leads to the simpler condition
\begin{align}
	\text{Re}( \varepsilon_j(k,\omega))=\omega \text{,}
	\label{Eq:condition}
\end{align}
which is often found to provide a satisfactory approximation. Here, the $\{\varepsilon_j\}$ denote the eigenvalues of the effective Hamiltonian (see Eq.~(\ref{eq:green1})), which are straightforward to compute. In fact condition (\ref{Eq:condition}) is often used in literature \cite{Stefanucci2010,Woelfle2018,Kozii2017}. Solving Eq.~(\ref{Eq:condition}) leads to $M$ curves $\omega_j^{QP}(k)$  which determine the complex poles $\{e_j(k)\}$ by:
\begin{align}
	e_j(k)  = \varepsilon_j(k,\omega_j^{QP}(k)) \text{.}
	\label{Eq:band}
\end{align}
The success of this approximation is intuitively clear since we are mainly interested in long-living excitations, which correspond to poles close to the real-frequency-axis \footnote{We note that the imaginary part in Eq.~(\ref{Eq:band}) is in general only proportional to the quasiparticle lifetime, and is rescaled by the quasiparticle weight $Z$ in common quasiparticle descriptions \cite{Stefanucci2010,Economou2006,Martin2016,Fabrizio2020}. Here, corroborated by our numerical results, assuming $Z \approx 1$ provides a satisfactory approximation.}. A selfenergy which only varies slowly with respect to the imaginary-frequency is then well approximated by their value along the real-frequency-axis. As expected by these arguments, our numerical results indicate that Eq.~(\ref{Eq:condition}) is less precise for fast decaying (short-lived) excitations.

\begin{figure}[htp]
	\centering
	\includegraphics[width=1.01\linewidth]{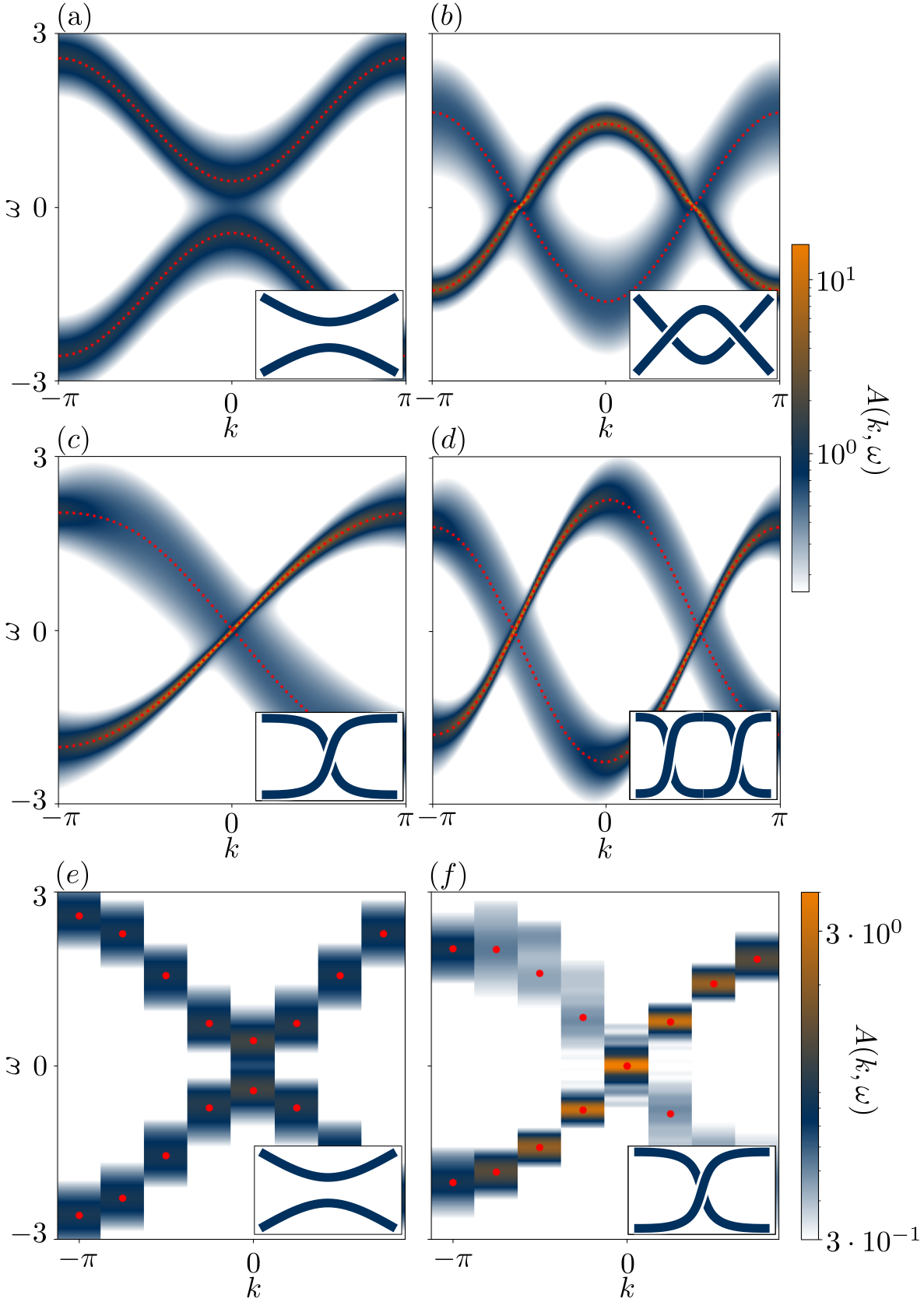}
	\caption{Spectral functions $A(k,\omega)$ in various braiding phases computed using NEGF ((a), $N = 200$ unit cells) or ED ((b), $N = 8$ unit cells). The corresponding link diagrams are shown in the insets. (a) and (e): trivial braid $c=0$,$\gamma=1$ with $J=1.5,\tau_0=1.0,\tau_1=0.0,v_0=0.0, v_1=0.0, U_a=1.5,U_b=0.1$ ; (b) trivial braid with crossings $c=0$,$\gamma=\sigma_1 \sigma_1^{-1}$ and parameters $J=1.5,\tau_0=1.0,\tau_1=0.0,v_0=0.0, v_1=0.0, U_a=1.5,U_b=0.1$; (c) lower left and (f) right: braid within phase $c=1$,$\gamma=\sigma_1 $ and $J=1.0,\tau_0=1.0,\tau_1=0.0,v_0=-1.0, v_1=0.0, U_a=1.5,U_b=0.1$; (d) braid in phase $c=2$,$\gamma=\sigma_1 \sigma_1 $  and $J=1.0,\tau_0=0.2,\tau_1=1.0,v_0=-0.2, v_1=-1.0, U_a=1.5,U_b=0.1$. The red dotted line shows the real-part of the quasiparticle dispersion as a guide to the eye.}
	\label{fig:knots}
\end{figure}
\subsection{\label{sec:CBG} Classification with Braid-Groups}
Since for ordinary interacting systems the band-structure can be understood in terms of a complex quasiparticle dispersion, previous insights on complex bands in NH Hamiltonians may be adapted to our present setting of quasiparticle spectra.
Specifically, in 1D it was found that a complex band structure $\{e(k)\}$ can be classified within the framework of braiding groups $\mathcal{B}_n$ \cite{Li2020,Wojcik2020,Hu2021,Wang2021,Rui2022,Zhong2023}. Given a concrete quasiparticle spectra with $M$ separable complex energies bands $e_n(k)$ depending on the momentum $k\in[-\pi,\pi]$, pictorially they build up $M$ strings in a solid torus spanned by the complex plane and the periodic Brillouin zone (see Fig.~\ref{fig:overview} (b)). These strings form different conjugacy classes of braids which cannot be continuously deformed into each other without band-touchings. Consequently, two NH band structures are topological distinct if they obey a different braid class. Using the $M-1$ Artin generators\cite{Zhong2023} $\{\sigma_j\}$ in a $M$-band system we can also denote each braid by a braid word $\gamma$. For example, the generator $\sigma_j$ interchanges the string $j$ with the string $j+1$, such that the first over-crosses the second, respectively the inverse $\sigma_j^{-1}$ describes the under-crossing. Generally, for a braid with $M$ strings, the braid word indicates whether two neighboring strings may over-cross or under-cross after being projected onto a plane. In our case, we project onto the real-part and flatten the imaginary-part (Fig.~\ref{fig:overview}(c)) as is physically motivated by notion of a spectral function. Topologically protected crossings then directly correspond to nodal spectral functions.

In the following we will concentrate only on two-band models, that is the simplest non-trivial braid group $\mathcal{B}_2$. It obeys only one Artin-generator $\sigma_1$ and is isomorphic to the integers $\mathds{Z}$. 
Given a dispersion $\{e_1(k),e_2(k)\}$ the braiding class $c \in \mathds{Z}$ is defined by a winding number\cite{Rui2022,Wang2021}:
\begin{align}
	c = -i \int_{-\pi}^{\pi} \frac{dk}{\pi } \partial_k \text{ln}(e_1(k)-e_2(k)) \text{.}
	\label{EQ:conjugacyc}
\end{align}

With these braid classification tools we investigate the quasiparticle dispersion $\{e_1(k), e_2(k)\}$ in 1D.

\paragraph*{Braiding Topology in Real Systems.}
Using NEGF and ED, we compute the rGF in the frequency-momentum domain in thermal equilibrium for the model described by Eq.~(\ref{EQ:Hamiltonian}) with PBC so as to obtain the spectral function and the quasiparticle dispersion.
In Fig.~\ref{fig:knots} the spectral functions $A(k,\omega)$ are presented for different parameters. Here, the real-part of the computed quasiparticle dispersion $\{e(k)\}$ corresponds well to the maxima of the band-structure (red dotted line). The same conclusion is true for the imaginary-part describing the broadening of the bands as showed in the appendix \ref{appendix:broadening}. Both these observations confirm that the analytical definition of the quasiparticle dispersion provides a physically reasonable framework.
Furthermore, we compute the braiding class $c$, as visualized by the corresponding link diagram (inset), showing that the chosen model can realize different topological braiding phases due to its sublattice dependent interaction.
As expected, the braiding class is already visible from the spectral function and thus of immediate experimental relevance. We note that the ED results are qualitatively inline with the NEGF approach and indicate that the braiding classification is already meaningful for systems of small size.

\section{\label{sec:CBT} Consequences of the Braid-Topology}
\subsection{\label{sec:TC}Topologically Protected Crossings}
The complex nature of the band-structure (see Eq.~(\ref{Eq:band})) has interesting implications regarding crossings in the spectral function. Specifically, such a crossing occurs if two real-part projections of the quasiparticle dispersion intersect, i.e. if $\text{Re}(e_1(k)) = \text{Re}(e_2(k))$ at some momentum $k$ (see Fig.~\ref{fig:knots}). Since the imaginary degree of freedom is in general gapped, crossings are stable, i.e. every small perturbations only renormalizes their position, but cannot gap out the nodal spectral function (see (b) in Fig.~\ref{fig:overview}). 

\paragraph*{Local Stability.} Locally in  frequency-momentum-space, the only way to gap out an isolated crossing is to overcome the gap in the imaginary-part, which necessarily leads to a degeneracy in the complex energies. This generically results in an EP \footnote{Commonly, since exceptional point degeneracies have a lower codimension than complex diabolic degeneracies, it is expected that gapping out a crossing leads to EPs close to the crossing.} where the rGF(see Eq.~(\ref{eq:green1})) becomes non-diagonalizable and two eigenvectors coalesce \cite{Bergholtz2019,Miri2019,Ashida2020,Kozii2017}.  
\begin{figure}[htp]
	\centering
	\includegraphics[width=1.0\linewidth]{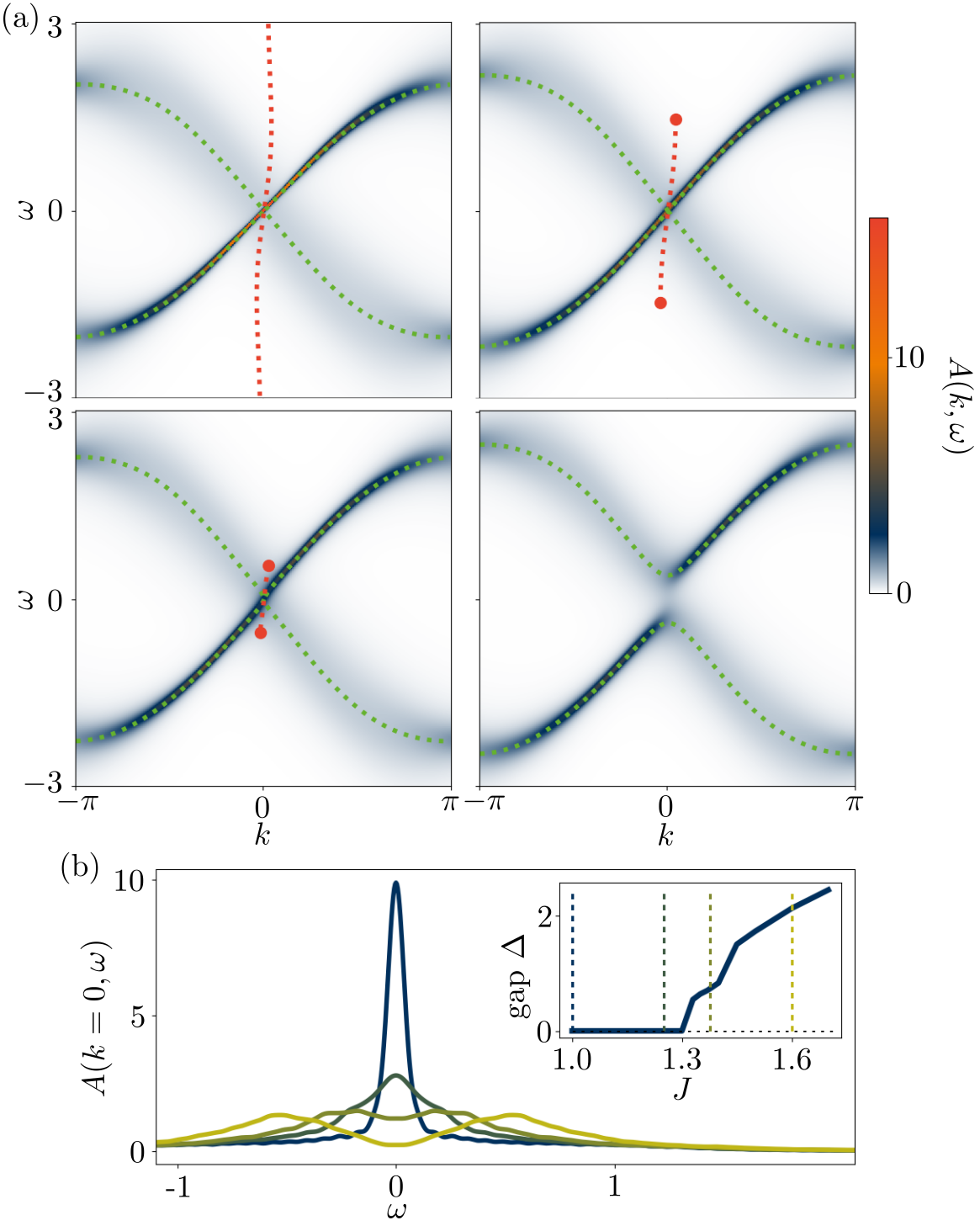}
	\caption{(a) Spectral function $A(k,\omega)$ for $J =\{1.00,1.15,1.26,1.45\}$ (panels in reading order) calculated using NEGF for $N=200$. Green dashed lines correspond to real-part of quasiparticle dispersion, in good agreement with intensity maxima. Red dashed lines mark imaginary Fermi arcs connecting the EPs (red dots). Initially ($J=1.0$) the system is in the braiding phase $c=1$, with increasing $J$ EPs move closer to the crossing, and finally merge resulting in a gapped trivial phase $c=0$.
	(b) Spectral function $A(k=0,\omega)$ at fixed momentum and $J =\{1.00,1.25,1.37,1.6\}$ (from blue to yellow), calculated within ED for $N=7$. Inset: gap $\Delta$ for different perturbation strengths $J$, calculated using the distance between the intensity maxima. Other parameters in all plots: $\tau_1 = 1.0, v = -1.0,U_a=1.5, U_b=0.1,T=5$, periodic boundary conditions}.
	\label{fig:gapping}
\end{figure}

Fig.~\ref{fig:gapping} (a) shows how a typical crossing locally gaps out, the results are computed using NEGF and the  perturbation is an increase in $J$ (see Eq.~(\ref{EQ:Hamiltonian})).  
First, for a wide range of perturbations the single nodal point shifts but stays intact (Fig.~\ref{fig:gapping} upper row). Finally, a pair of EPs approaches and as both merge, the crossing gaps out(Fig.~\ref{fig:gapping} lower row). The occurrence of EPs close to a phase transition is a robust feature as well as the so called imaginary Fermi-arc (red dashed line) \cite{Bergholtz2019,Kozii2017} which, though spectroscopically invisible, facilitates a complex analytical understanding of the process. Specifically, the imaginary Fermi-arc intersects the crossing and thus connects the two EPs. Since ED is limited to small systems sizes, the rGF there lacks a sufficient resolution in momentum-space for locating the EPs. However, the stable crossing may be visualized by the spectral function for different perturbations at momentum $k=0$ (Fig.~\ref{fig:gapping} (b)). In agreement with the NEGF data on larger systems, a splitting of a single peak in two smaller ones can be seen, which occurs only at significant perturbation strengths $J$ (Fig.~\ref{fig:gapping} (b) inset). This observation also complements earlier findings of EPs close to the Kondo transition in \cite{Michishita2020,Peter2021}.

\paragraph*{Global Stability.} A second way to gap out a crossing is by including a second nodal point, and then smoothly deforming the system until both may annihilate each other. As an example, the braid $\gamma=\sigma_1 \sigma_1^{-1} = 1$ (Fig.~\ref{fig:knots} upper right) can be smoothly deformed without EPs into an insulator (Fig.~\ref{fig:knots} upper left). In this scenario, a local crossing can get continuously annihilated by its inverse without the necessity of an EP transition.
  
These phenomena can be generalized to the full Brillouin zone, respectively the "global" conjugacy class $c$ (Eq.~(\ref{EQ:conjugacyc})) induces at least $\vert c \vert$ crossings which can only be gapped by EPs, thus for a non-trivial dispersion like the lower ones in Fig.~\ref{fig:knots} the crossings are also globally stable. Eventually, stable semimetallic phases are the first consequence of the previously introduced classification in braiding groups.

Finally, it is worth to mention that the stability of nodal points contradicts with a Hermitian description of quasiparticle bands, where already a small perturbation can destroy a crossing. Further, since this effect is induced by the non-Hermiticity of the selfenergy, we may argue that crossings of electronic band-structures are genuinely stabilized by interactions with a non-trivial orbital structure, in our case a sublattice dependence.

\subsection{\label{sec:NRSPE} Non-Reciprocal Single-Particle Green's Function Excitations}
\begin{figure}[htp]
	\centering
	\includegraphics[width=1.0\linewidth]{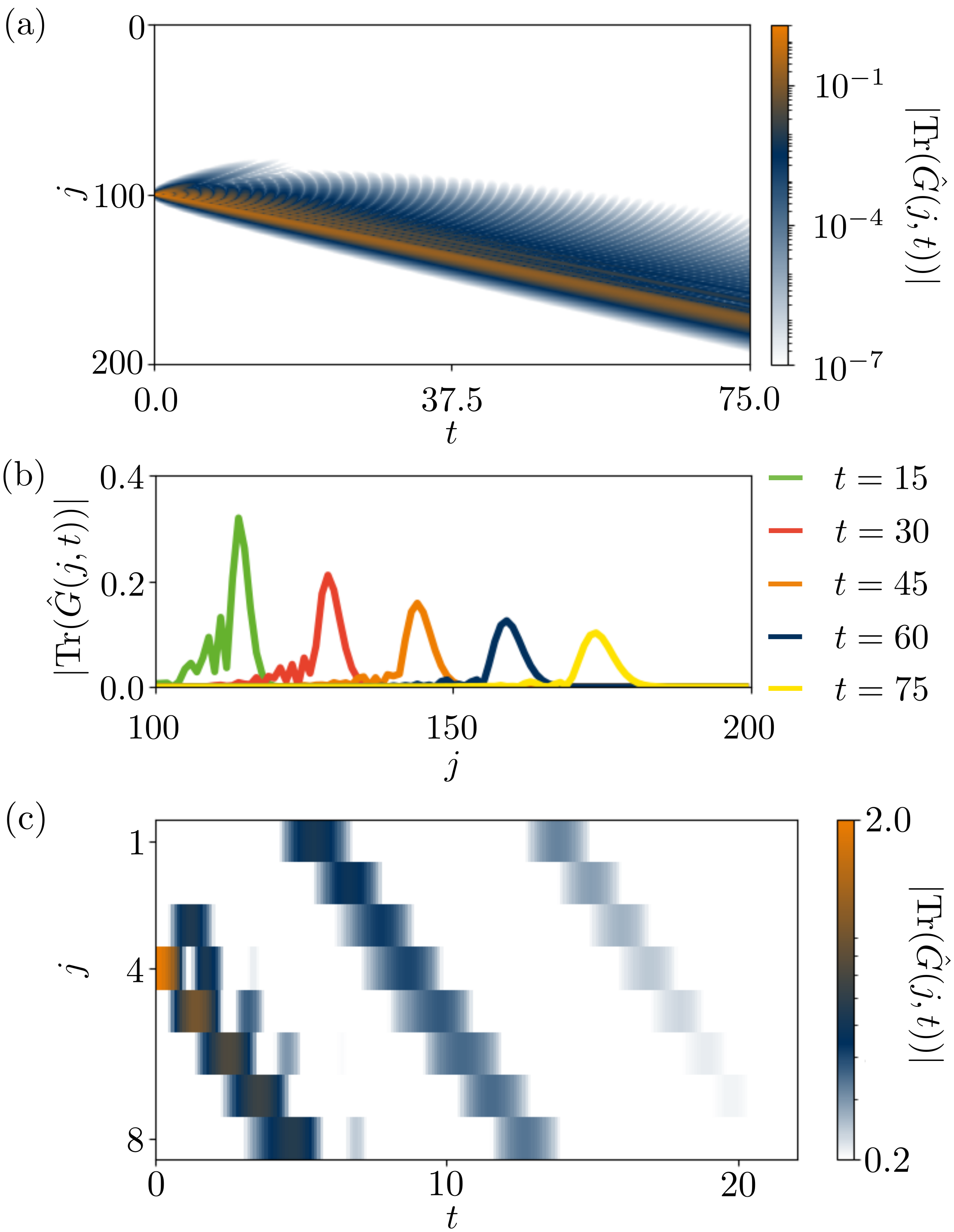}
	\caption{(a) and (c): Absolute value of the trace $\lvert \text{Tr}(\hat{G}(j,t) )\rvert$ in real-space within the braiding phase $c=1$ and $J=1.0,\tau_0=1.0,\tau_1=0.0,v_0=-1.0, v_1=0.0, U_a=1.5,U_b=0.1, N=200 (\text{NEGF}), N=8 (\text{ED})$, showing a non-reciprocal behaviour. (b): Absolute value for fixed times $t$ (same parameters as in (a)), with a Gaussian ballistically moving peak. (c): PBC and small system size leading to circulating behaviour.}
	\label{fig:drift_G}
\end{figure}

Besides the quasiparticle energy, the group velocity $v_j(k)$ corresponding to an excitations is of interest, which we defined by the momentum derivative of the quasiparticle energy Eq.~(\ref{Eq:band}):
\begin{align}
	v_j(k)=\partial_k \text{Re}(e_j(k)) \text{.}
\end{align}
The velocity $v_j(k)$ in combination with the corresponding inverse lifetime $\text{Im}(e_j(k))$ allows to gain insights at a qualitative level on the propagation dynamics of single-particle excitations and the timescales for which certain modes are relevant.
\paragraph*{Non-Reciprocal Crossings}
An ordinary crossing often obeys group velocities in opposite directions, i.e. $\text{sgn}(v_1(k)) =- \text{sgn}(v_2(k))$, and a stable crossing typically consists of a broader and a sharper band, as their lifetimes differ due to the imaginary gap. Then, the excitations of one band decay slower than the excitations of the second band that move in the opposite direction. Consequently, a perturbation that is equally exciting left- and right-movers close to the crossing, is expected to exhibit non-reciprocal dynamics. In the spectral function $A(k,\omega)$ Fig.~\ref{fig:knots} these local non-reciprocal crossings are very well visible. By looking at the rGF in real-space and real-time, we see that the quasiparticle dynamics indeed acquires a non-reciprocal asymmetry very fast in the non-trivial braiding phase (see Fig.~\ref{fig:drift_G}).  

Observing this local non-reciprocity requires selectively exciting quasiparticles close to a crossing. By contrast, in the upper right band-structure in Fig.~\ref{fig:knots}, the second crossing counters the effect induced by the first crossing, thus preventing clear non-reciprocal effects in real-space.
A more stable approach to reveal the non-reciprocal dynamics is to look at a regime, where the longest-living excitations are close to a crossing. This ensures that on intermediate and larger time-scales these excitations dominantly determine the dynamics.

Generalizing non-reciprocity to a global topological perspective, we find that every odd braiding number $c=2n-1$ ($n\in\mathds{Z}$) exhibits non-reciprocal behavior in the above dynamical sense unless the dispersion is non-analytic (see proof in Appendix \ref{appendix:proof_nr}). Alternatively, defining reciprocity as equivalent to inversion symmetry $k\rightarrow-k$ (as in \cite{Zhong2023}), it follows that every non-trivial braid is non-reciprocal. Apart from possible fine-tuned exceptions, we find that any non-trivial braidings are at least good candidates to also see the previously described local non-reciprocity.

\begin{figure}[htp]
	\centering
	\includegraphics[width=1.0\linewidth]{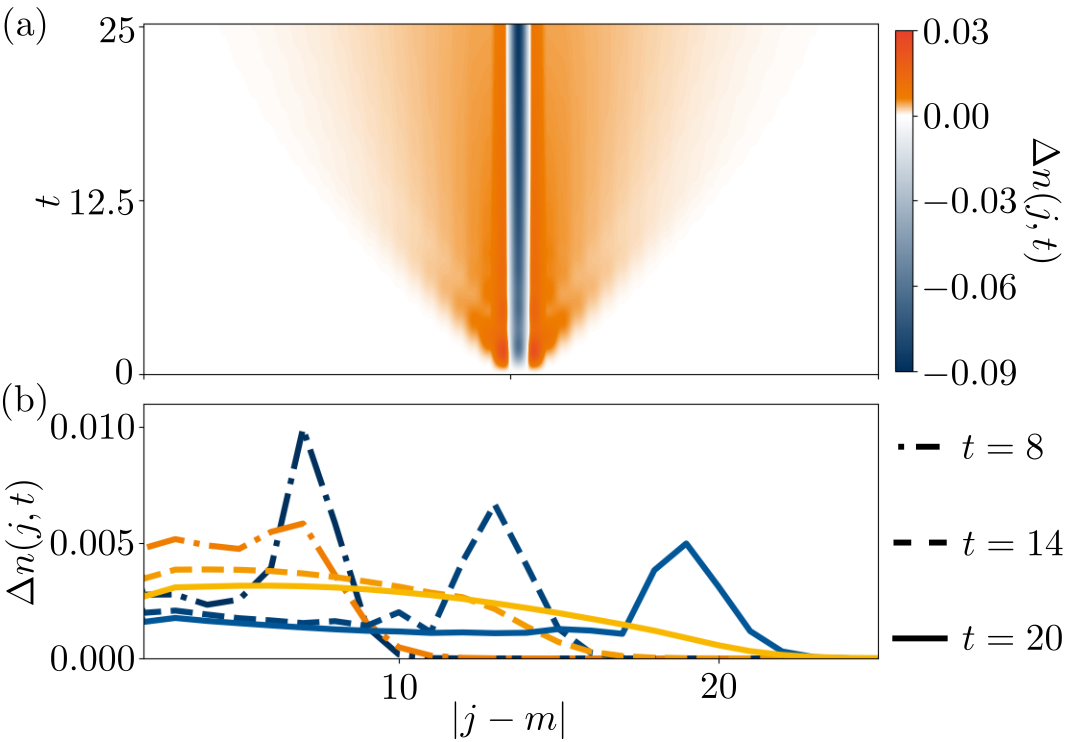}
	\caption{Time dependent change of the density $\Delta n(j,t)$ illustrating charge propagation after an onsite quench by switching on the a potential $V_m \left(n_A(m)+n_B(m)\right)$ at site $m=25$($V_m=1.0$): (a) reciprocal system with trivial braiding and parameters $J=1.5,\tau_0=1.0,\tau_1=0.0,v_0=0.0, v_1=0.0, U_a=1.5,U_b=0.1$, (b) real-space cuts of $\Delta n(j,t)$  as a function of distance $\vert j-m \vert$ to the perturbed site $m$ for fixed times $t$ in the non-reciprocal braiding phase $c=1$ (cf.~Fig.~\ref{fig:overview}(d)), right-moving part shown in blue and the left-moving part in orange. Parameters are $J=1.0,\tau_0=0.0,\tau_1=1.0,v_0=0.0, v_1=-1.0, U_a=1.5,U_b=0.1$.}
	\label{fig:charge_propagation}
\end{figure}

\subsection{\label{sec:NRCP}Non-Reciprocal Charge Propagation}
In addition to the previously described non-reciprocal rGF properties, non-reciprocity in non-trivial braiding phases can be observed in non-equilibrium dynamics of the charge density after a local parameter quench, indicating that non-reciprocal quasiparticles may influence common measurable quantities. 
\paragraph*{Quench Setup}
To this end, we assume that the system is initially prepared in thermal equilibrium at temperature $T=5$, before an on-site potential  $H_{t>0}=H+V_j \left(n_{j,A}+n_{j,B}\right)$ at site $j$ is switched on at $t=0$. 
Pictorially speaking, the potential pushes particles away from the chosen site $j$, and creates a charge perturbation spreading over the whole system, which is sensitive to the quasiparticles described by the rGF. A similar quench setup has been used to study non-equilibrium dynamics in spin-$\frac{1}{2}$ XXZ-Chains and Fermi-Hubbard models \cite{Karrasch2016,Karrasch2017,Karrasch2014,Bertini2021}.
The propagation of the density excitation can be seen in the particle-density per site $\Delta n(j,t) = \sum_{\alpha=\{A,B\}} \langle n_{j,\alpha}(t) - n_{j,\alpha}(0)\rangle $.
Here $n_{j,\alpha}(t)$ denotes the density operator in the Heisenberg picture.
The particle-density after the quench was computed using NEGFs for the insulator model with trivial braiding and the non-trivial model with conjugacy class $c=1$ of the previous section within open boundary conditions and $N=50$ unit cells.
\paragraph*{Non-Reciprocity in the Topological Phase} 
As shown in Fig.~\ref{fig:overview} (d) and  Fig.~\ref{fig:charge_propagation} (a) the charge perturbation spreads over the system in time, building-up a light-cone structure.
Interestingly, this light-cone is now asymmetric for a non-reciprocal band-structure, i.e. the left and right moving excitation both propagate ballistically for short times, but with increasing time the former spreads diffusively, while the latter still propagates linearly.
This picture is in line with the lifetime difference of the right- and left-moving quasiparticles in the spectral function, suggesting that the non-reciprocity of the quasiparticle bands consequently leads to non-reciprocity in the real-space charge dynamics. The non-reciprocity gets even more visible by looking at real-space cuts of $\Delta n(j,t)$ for fixed times $t$(Fig.~\ref{fig:charge_propagation} (b)), where the excitation moving to the right shows a stable ballistic peak structure (blue), contrary to the smoothly broadening front moving to the left (orange).
For comparison, we also simulated the reciprocal system in Fig.~\ref{fig:charge_propagation} (a) which is found to exhibit a symmetric charge propagation as expected.  This clarifies that the non-reciprocity is indeed an intrinsic effect of the system and does not result from the shape of the perturbation or the open boundary conditions.
The asymmetric charge-propagation exemplifies that the non-trivial braiding phases have observable effects beyond the spectral function $A(k,\omega)$, thus highlighting the relevance of NH topology for the dynamics of interacting many-body systems. 
\subsection{\label{sec:LEUQ}Pseudo-Chiral Mode Limit}
The lifetime difference at a crossing may be tuned up to a point, where one mode acquires (practically) infinite lifetime. In our model, this limit is realized by switching off the interactions on the $B$-sublattice ($U_B=0$). For strongly sublattice dependent interactions (compare to Eq.~(\ref{EQ:Hamiltonian})), interactions on one sublattice do not necessarily affect quasiparticles on the other sublattice. This can be illustrated by the inverse rGF in matrix representation (see also Eq.~(\ref{eq:green1})):
\begin{align}
	\label{EQ:retardedG}
	\hat{G}^R(k,\omega)^{-1} = \omega\mathds{1} - \begin{pmatrix}
		h_{AA}(k) + \Sigma(k,\omega) & h_{BA}^{\ast}(k) \\
		h_{BA}(k) & h_{BB}(k)
	\end{pmatrix}.
\end{align}
Here the non-interacting Bloch-Hamiltonian $\hat{h}(k)$ was used in its matrix components and the self-energy $\Sigma(k,\omega)$ is only non-zero on the $AA$-entry, which is a consequence of $U_B = 0$ in Eq.~(\ref{EQ:Hamiltonian}).
In case of a vanishing coupling $h_{BA}(k^{QP})=0$ at some momentum $k^{QP}$, the effective Hamiltonian is diagonal. 
The corresponding eigenvalues are then easily obtained by $\varepsilon_1(k^{QP},\omega)= h_{AA}(k^{QP}) + \Sigma(k^{QP},\omega)$ and $\varepsilon_2(k^{QP},\omega)= h_{BB}(k^{QP})$. It follows immediately that the inverse lifetime $\text{Im}(\varepsilon_2(k^{QP},\omega))=0$ of the second eigenvalue vanishes. Consequently, the spectral function on the $B$-sublattice  $A_{B}(k^{QP},\omega) = -\frac{1}{\pi} \text{Im}\left( G^R_{BB}(k^{QP},\omega) \right) \propto \delta(\omega - h_{BB}(k^{QP}))$ gets a sharp peak at $\omega=h_{BB}(k^{QP})$(see Appendix \ref{appendix:exact_delta}), also visible in the full spectral function. In the vicinity of $k^{QP}$ the coupling might be still small, leading to a sharp band dispersion \footnote{For small couplings $h_{BA}(k^{QP})\ll1$ this picture is still a good approximation, and the sharp band characteristic stays intact (see Fig.~\ref{fig:knots} lower row).}.  Interestingly, these modes do not preclude the non-integrability of the model (see section \ref{sec:NI}), and thus constitute individual ballistic modes embedded in a fully chaotic environment.

The corresponding quasiparticles in the vicinity of $k^{QP}$ obey approximately an infinite lifetime and have energy
\begin{align}
	\label{EQ:energy_NIM}
	e^{QP}(k)\approx v^{QP}(k-k^{QP}) + h_{BB}(k^{QP}),
\end{align}
where $v^{QP}$ corresponds to the group velocity $v^{QP} =\partial_k h_{BB}(k)\vert_{k^{QP}}$. Such excitations have a strongly pseudo-chiral character, in particular if the counter-propagating excitation has a finite (short) lifetime. This limit of non-interacting modes provides an intuitive picture for the non-reciprocal effects in our model. Further, in the extreme limit of practically uni-directional motion it constitutes a NH one-dimensional counterpart of chiral edge states familiar from two-dimensional quantum Hall systems. However, the non-reciprocity of crossings described above is much more general and its visibility in observables extends far beyond the parameter regime of the pseudo-chiral limit.

\section{\label{sec:CON} Concluding discussion}
With this work we propose the notion of nodal spectral functions, i.e. spectral functions that exhibit robust crossings in the real part of the dispersion that are stabilized by the NH topology of complex quasiparticle bands, in particular by their braid class. We have exemplified the generic occurrence of such nodal NH topological phases in correlated 1D systems by numerically solving a two-banded microscopic lattice model with a sub-lattice dependent interaction strength. Remarkably, this simple model system already hosts a variety of distinct braiding phases, that leave clear fingerprints both in the spectral function and in non-reciprocal charge dynamics. At the level of NH tight-binding models, the topological classification of complex energy bands in terms of braiding classes has recently been discussed in several studies \cite{Wang2021,Li2021,Rui2022,Hu2021,Wojcik2020,Nehra2022,Koenig2023}. Our present work establishes spectral functions of correlated electron systems as a natural physical setting where these concepts enter the stage to explain the abundance and robustness of nodal band structures. Exceptional points, the main focus of previous work on nodal NH topology, here hallmark the transitions between different braiding phases. Our insights complement an earlier result \cite{Michishita22020}, where EPs close to a Kondo transition were discussed.

The relation between nodal spectral functions and non-reciprocal dynamics or transport may be intuitively understood from the generically finite (except at an EP) gap in the imaginary part of the quasiparticle spectrum right at the crossing of the real part. This amounts to a  lifetime difference for excitations close to the nodal point, i.e. the left-mover and right-mover exhibit a different degree of ballistic coherence. As a consequence, we have identified non-reciprocal dynamics both in the real-space GF (see Fig.~\ref{fig:drift_G}) and in the charge propagation after a local quench (see Fig.~\ref{fig:charge_propagation}). In the context of disordered systems with sub-lattice dependent impurity scattering \cite{Michen2022}, we note that a related non-reciprocity in mesoscopic quantum transport has also been explained by topological properties of an effective NH Hamiltonian.

Our model contains an interesting phenomenology in the limit of an entirely sub-lattice selective interaction ($U_B =0$). There, while we find that the model remains non-integrable, an isolated free (or perfectly ballistic) mode emerges at the protected level crossing. This situation is reminiscent of the uni-directional transport in a chiral edge state known from two-dimensional quantum Hall systems. While fermion doubling (or in 1D simply Bloch's theorem) seems to exclude a uni-directional mode in 1D, a sizable splitting in the imaginary part may be seen as a non-reciprocal damping of ballistic transport, thus burying the counter-propagating mode deeply in the lower complex half-plane and largely blurring its visibility in the spectral function, respectively. To understand this phenomenon in more depth, we have presented an analytical approach to the non-interacting excitation, thus providing a complementary microscopic view of the non-reciprocal dynamics in this extreme limit. However, we note that the occurrence of non-reciprocity from real-part crossings accompanied by an imaginary gap is far more general than the $U_B =0$ limit.

From an experimental perspective, the observation of nodal spectral functions and their distinction from gapped systems is conceptually simple but in practice strongly depends on the ability to tune and resolve the lifetime difference between orbitals or sub-lattices. The predicted non-reciprocal dynamical effects additionally rely on the ability to selectively excite modes close to nodal points.  

\begin{acknowledgments}
	We acknowledge financial support from the German Research Foundation (DFG) through the Collaborative Research Centre SFB 1143 (Project-ID 247310070), the Cluster of Excellence ct.qmat (Project-ID 390858490). Our numerical calculations have been performed at the Center for Information Services and High Performance Computing (ZIH) at TU Dresden.
\end{acknowledgments}

\appendix

\section{Analytical Bandwidth}
\label{appendix:broadening}
\begin{figure}[htp]
	\centering
	\includegraphics[width=1.0\linewidth]{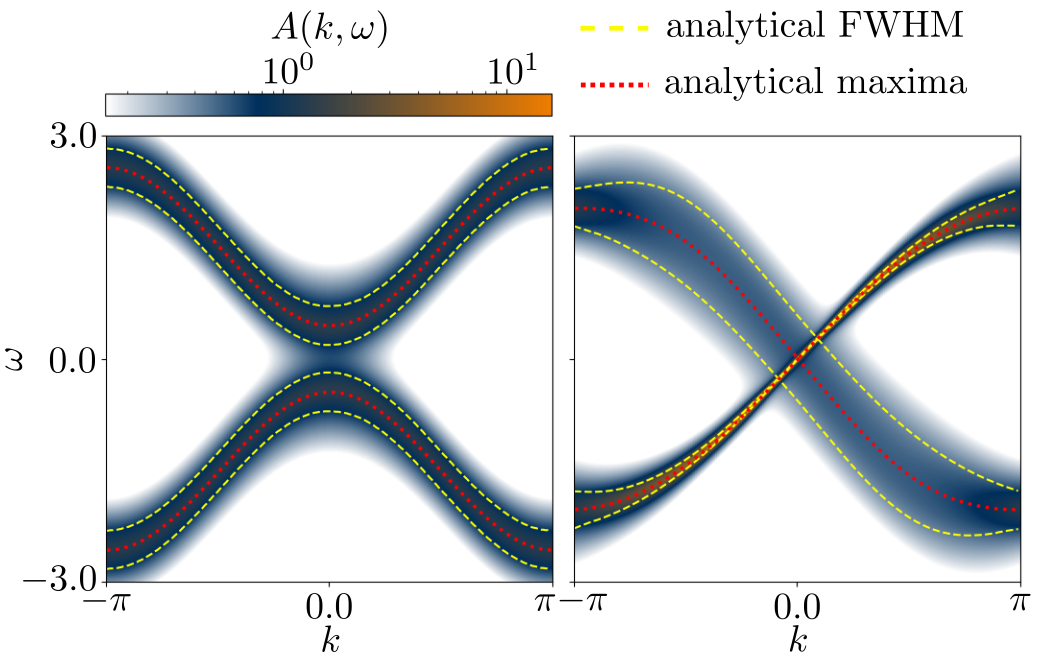}
	\caption{Spectral functions $A(k,\omega)$ and the corresponding description using the quasiparticle approximation $\{e_j(k)\}$. Left: trivial braid with $J=1.5,\tau_0=1.0,\tau_1=0.0,v_0=0.0, v_1=0.0, U_a=1.5,U_b=0.1, N=200$. Right: braid within phase $c=1$ and $J=1.0,\tau_0=1.0,\tau_1=0.0,v_0=-1.0, v_1=0.0, U_a=1.5,U_b=0.1, N=200$. $\text{Re}(e_j(k))$ describes the maxima(red dotted line), while $\text{Im}(e_j(k))$ is an approximation for the full-width-at-half-maximum(FWHM) ,here $\text{Re}(e_j(k)) \pm \text{Im}(e_j(k))$ is represented by yellow dashed line. In total we see a very good agreement between the analytical FWHM and the broadening of the bands.}
	\label{fig:bandwidths}
\end{figure}

\newpage
\section{Winding using ED}

\label{appendix:ED_winding}
\begin{figure}[htp]
	\centering
	\includegraphics[width=1.0\linewidth]{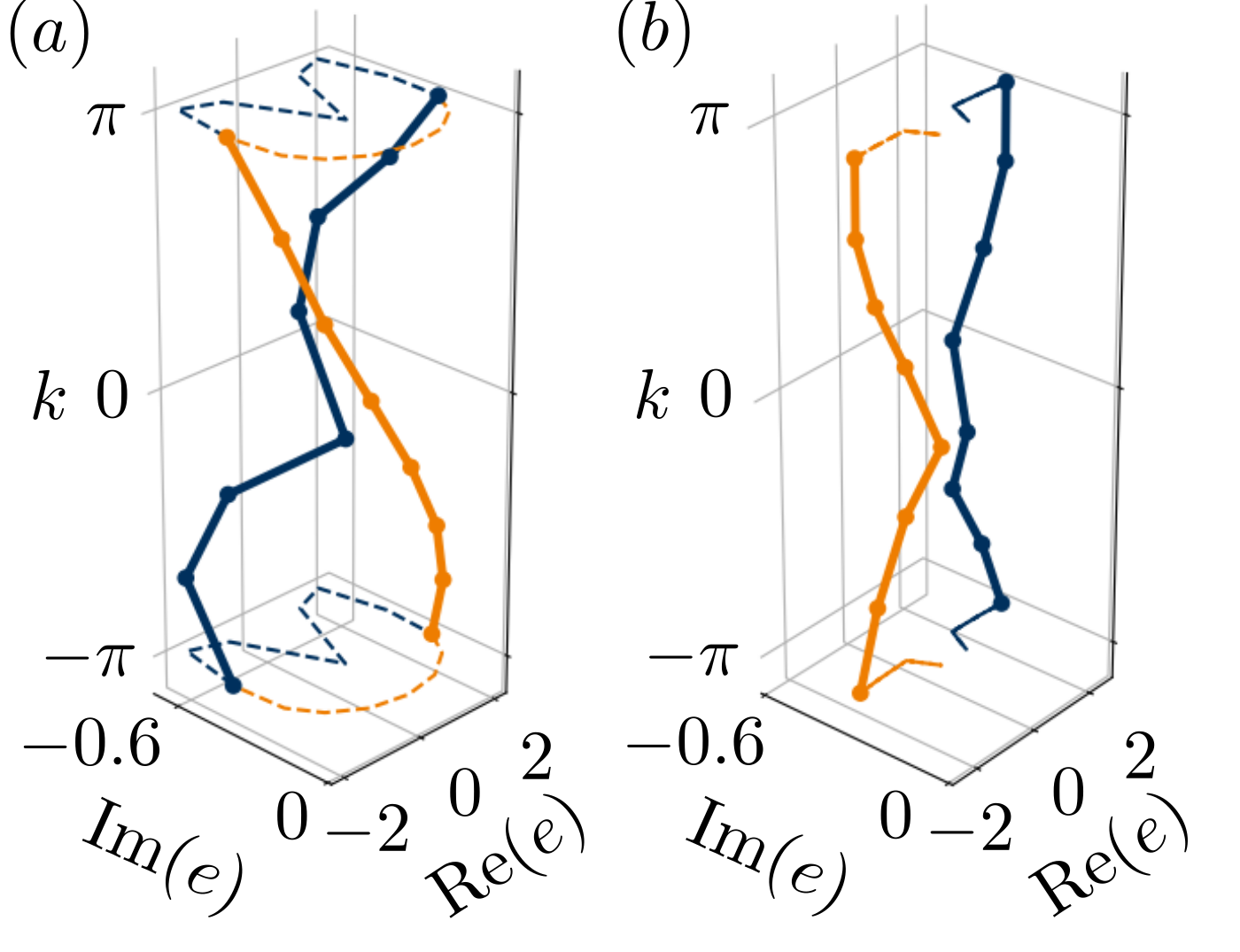}
	\caption{Winding of the complex dispersion $\{e_j(k)\}$ calculated with ED. (b): trivial braid with $J=1.5,\tau_0=1.0,\tau_1=0.0,v_0=0.0, v_1=0.0, U_a=1.5,U_b=0.1, N=14$. (a): braid within phase $c=1$ and $J=1.0,\tau_0=1.0,\tau_1=0.0,v_0=-1.0, v_1=0.0, U_a=1.5,U_b=0.1, N=14$. Despite the limited system-size a clear braiding characteristic is visible.}
	\label{fig:winding_ED}
\end{figure}

\section{Equivalence of the Reciprocity of the Bandstructure and the Refelection symmetry}
\label{appendix:proof_nr}
Given a 1D complex dispersion $\{e_j(k)\}$ as function of the momentum $k$ in the Brillouin zone consisting of $M$ bands, we assume that such a dispersion is reciprocal if for every mode $e_j(k_1)$ at momentum $k_1$ exist a mode $e_l(k_2)$ at momentum $k_2$ with the same complex energy $e_j(k_1) = e_l(k_2)$,i.e. lifetime and excitation energy are the same, but both move in opposite direction, thus their group velocities are equal up to a minus sign $\partial_k \text{Re} (e_j(k))\vert_{k_1} = -\partial_k \text{Re} (e_l(k))\vert_{k_2}$. Notice that the band index $j,l$ can be equal or different. 
We will now show that for uneven braids in the braiding group $\mathcal{B}_2$ assuming both bands are analytic(real and imaginary part can be expanded in a Taylor series), the reciprocity condition above is equivalent to the assumption that the dispersion is reflection symmetric $k\rightarrow-k$ after applying a momentum translation in the periodic Brillouin zone.
This leads to a contradiction since a non-trivial braiding cannot obey reflection symmetry, from what follows uneven braidings are incompatible with reciprocity. 

\begin{figure}[htp]
	\centering
	\includegraphics[width=1.0\linewidth]{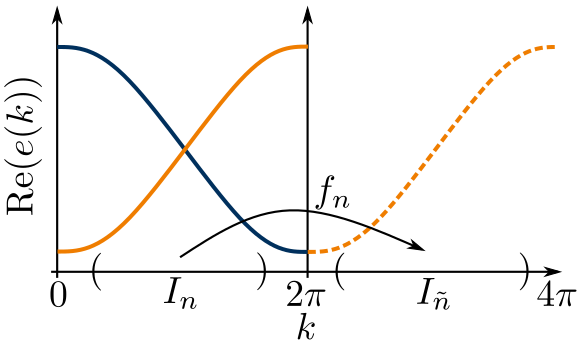}
	\caption{Unfolding procedure for braidings with uneven braiding $c$, by doubling the Brillouin zone one gets an effective one band description. Schematic illustration of the partition in intervals $\{I_n\}$ on which one can define the smooth shift functions $f_n$.}
	\label{fig:scheme_proof}
\end{figure}

We start by unfolding the braid, doubling the Brillouin zone and see the original braid as one complex band $e(k)=e_1(k)\oplus e_2(k)$ in the expanded Brillouin zone, this is possible since we assume an uneven braid(Fig.~\ref{fig:scheme_proof}).
Consequently $e(k)$ is periodic on the interval $k\in[0,4\pi]$ and bounded in its complex energy. From now on we concentrate on the real-part $\text{Re} (e(k))=e_{\text{R}}(k)$, which also fulfils these conditions.
Here $e_{\text{R}}(k)$ can divided in an even number of pairwise disjunct intervals $\{I_n\}$, where each $I_n$ has a partner $I_{\tilde{n}}$ with a function $f_n(k)\in I_{\tilde{n}}$ such that $e_{\text{R}}(k)=e_{\text{R}}(f_n(k))$ for $k\in I_n$.
Given an interval $I_n$ we choose now the partner interval $I_{\tilde{n}}$ such that $\partial_k e_{\text{R}}(k)\vert_{k_1} = -\partial_k e_{\text{R}}(k)\vert_{f_n(k_1)}$, what defines a differential equation which determines $f_n(k)$.
\begin{align}
	\partial_k e_{\text{R}}(f_n(k)) \vert_{k_1} &= \partial_k e_{\text{R}}(k)\vert_{f_n(k_1)} \partial_k f_n(k)\vert_{k_1} \\
	&= -\partial_k e_{\text{R}}(k)\vert_{k_1} \partial_k f_n(k)\vert_{k_1} \\
	\partial_k e_{\text{R}}(f_n(k)) \vert_{k_1} &= \partial_k e_{\text{R}}(k) \vert_{k_1} 
\end{align}
Subtracting the second from the third line leads an equation for $f_n(k)$.
\begin{align}
	\partial_k f_n(k) &= -1 \\
	f_n(k) &= -k + c_n
\end{align}
After solving the differential equation we see that the $f_n(k)$ just consist on a momentum inversion and a fixed shift $c_n$.
Thus for each interval pair $I_{\tilde{n}},I_n$ the band fulfils a local reflection symmetry $e_{\text{R}}(k)=e_{\text{R}}(-k + c_n)$.

In the next step we show that under the assumption that $e_{\text{R}}(k)$ is analytic everywhere, each local reflection symmetry is not only valid on the intervals $I_{\tilde{n}},I_n$ but also on the whole Brillouin zone.
For that we expand $e_{\text{R}}(k)$ at $k_1\in I_n$ and at $c_n - k_1\in I_{\tilde{n}}$:
\begin{align}
	e_{\text{R}}(k) \approx& e_{\text{R}}(k_1) + \partial_k e_{\text{R}}(k)\vert_{k_1} (k-k_1) \nonumber \\ &+ \frac{1}{2} \partial_k^2 e_{\text{R}}(k)\vert_{k_1} (k-k_1)^2 ... \\
	e_{\text{R}}(k) \approx& e_{\text{R}}(-k_1 + c_n) + \partial_k e_{\text{R}}(k)\vert_{-k_1 + c_n} (k+k_1 - c_n) \nonumber \\ &+ \frac{1}{2} \partial_k^2 e_{\text{R}}(k)\vert_{-k_1 + c_n} (k+k_1 - c_n)^2 ...\\
	=& e_{\text{R}}(k_1) - \partial_k e_{\text{R}}(k)\vert_{k_1} (k+k_1 - c_n) \nonumber \\ &+ \frac{1}{2} \partial_k^2 e_{\text{R}}(k)\vert_{k_1} (k+k_1 - c_n)^2 ...
\end{align}
The replacement of the derivatives at the $ c_n - k_1$ follows from the reflection symmetry.
Now we using the property of analytic functions, that knowing the Taylor-expansion at one point necessarily determines the full dispersion  $e_{\text{R}}(k)$ on the Brillouin zone, so we now shift our function by $k \rightarrow k_s+\frac{c_n}{2}$ and sum up both expansions to gain the expansion at $k=\frac{c_n}{2}$ or in the new shifted space at $k_s=0$, it follows that all uneven multiples in $k_s$ vanish.
\begin{align}
	e_{\text{R}}(k_s) \approx C_0 + 0 \cdot k_s + C_2 k_s^2 + 0 \cdot k_s^3 + C_4 k_s^4 ...	
\end{align}
This expression is reflection symmetric around $k_s=0$.  
Consequently it follows that each reflection symmetry is simultaneously fulfilled by the dispersion $e_{\text{R}}(k)$ on the whole Brillouin zone, thus we conclude $e_{\text{R}}(k)$ obeys at least one reflection symmetry. 
Since we assumed in the beginning $e_j(k) = e_l(\tilde{k})$, i.e. if the real-part coincidence also the imaginary-part coincidence, thus also the imaginary-part has the same symmetry and so the full dispersion $e(k)$ is symmetric under reflection after a potential shift in momentum.
In the last step we backfold our dispersion onto the original Brillouin zone, however this operation does not destroy the symmetry.

\section{Exact non-interacting mode in the spectral function} \label{appendix:exact_delta}

We start from Eq.~(\ref{EQ:retardedG}) and remember the  infinitesimal $i\eta$ in the $BB$-entry, we can neglect it in most cases for a non-trivial imaginary selfenergy $\Sigma(k,\omega)$. Here also the coupling $h_{AB}(k^{QP})=0$ vanishes so the rGF on the $B$-sublattice has no imaginary selfenergy, thus we can not neglect the $i\eta$. Being at momentum $k^{QP}$ where the coupling $h_{AB}(k^{QP})=0$ vanish, we get: 
\begin{align}	
	&\hat{G}^R(k^{QP},\omega)^{-1} = \omega\mathds{1} - \begin{pmatrix}
		h_{AA} + \Sigma & 0 \\
		0 & h_{BB} - i\eta
	\end{pmatrix}.
\end{align}
For the readability we omitting the arguments in the matrix entries, $h_{AA}(k^{QP}) \widehat{=} h_{AA}$, $\Sigma(k^{QP},\omega) \widehat{=} \Sigma$.
Due to the diagonal structure the spectral function $A(k^{QP},\omega)=-\frac{1}{\pi}\text{Im}(G^R(k^{QP},\omega))$ can now easily evaluated: 
\begin{align}	
	&\hat{A}(k^{QP},\omega) =\ \nonumber \\
	&-\frac{1}{\pi}\text{Im} \begin{pmatrix}
		(\omega - h_{AA} - \Sigma)^{-1} & 0 \\
		0 &(\omega -h_{BB} + i\eta)^{-1}
	\end{pmatrix}.
\end{align}
The spectral function on the $B$-sublattice is now defined as $A_{BB}(k^{QP},\omega)=-\frac{1}{\pi}\text{Im}(\omega -h_{BB}(k^{QP}) + i\eta)^{-1}$.
Taking the limit $\eta\rightarrow 0$ gives:

\begin{align}
	A_{BB}(k^{QP},\omega)&=\lim_{\eta\rightarrow0} \frac{1}{\pi}\frac{\eta}{(\omega -h_{BB}(k^{QP}))^2 + \eta^2} \nonumber \\ &=  \delta(\omega -h_{BB}(k^{QP})) 
\end{align}

As consequence the full spectral function $Tr(\hat{A}((k^{QP},\omega))) = A_{AA}(k^{QP},\omega) + A_{BB}(k^{QP},\omega)$ also shows a delta function.

\end{document}